# Towards a Flexible Framework for Algorithmic Fairness

Philipp Hacker[1], Emil Wiedemann[2] and Meike Zehlike[3]

**Abstract:** Increasingly, scholars seek to integrate legal and technological insights to combat bias in AI systems. In recent years, many different definitions for ensuring non-discrimination in algorithmic decision systems have been put forward. In this paper, we first briefly describe the EU law framework covering cases of algorithmic discrimination. Second, we present an algorithm that harnesses optimal transport to provide a flexible framework to interpolate between different fairness definitions. Third, we show that important normative and legal challenges remain for the implementation of algorithmic fairness interventions in real-world scenarios. Overall, the paper seeks to contribute to the quest for flexible technical frameworks that can be adapted to varying legal and normative fairness constraints.

## 1 Introduction

Discrimination by machine learning applications has emerged as a major challenge for the widespread deployment of AI technology. In recent years, it has not only sparked a vivid academic debate in computer science and law [DH+12; FF+15; BS16; Ha18; ZH+20; WM+20; He20], but has also garnered significant attention from the media and policymakers. Importantly, bias in AI systems does poses not merely a theoretical problem; rather, it is documented in an increasing number of reports and empirical studies. For example, Amazon had to discard its AI recruitment tool because it persistently discriminated against women [Re18]. Similarly, a study recently found that the machine learned model managing health populations in US hospitals exhibited racial bias [OP+19]. In fact, language itself encodes potentially discriminatory patterns, which surface in AI-based translation applications [CB+17]. Hence, algorithmic discrimination remains a pressing issue for both technology and policy, despite the research efforts of the past years.

Against this background, this short overview paper will approach the subject in three steps: first, it briefly covers core legal areas applying to algorithmic discrimination under EU law. Second, it explains how different technical strategies for remedying algorithmic discrimination can be combined using algorithmic fairness metrics and optimal transport. Third, it highlights that important normative challenges, both ethical and legal, remain for implementing technical solutions to bias in AI systems. The paper chiefly draws on longer articles in which the described algorithm, its functioning and legal implications were explained in greater detail [Ha18; ZH+20].

## 2 The Law of Algorithmic Discrimination

In EU law, there are two main fields that deal with discrimination in machine learning contexts: antidiscrimination law and data protection law. Both areas are tightly regulated at the EU level. The following paragraphs can only present a succinct summary of the application of that body of law to algorithmic discrimination.

---

[1] European University Viadrina, European New School of Digital Studies, Chair for Law and Ethics of the Digital Society, Große Scharrnstraße 59, 15230 Frankfurt (Oder), Germany, hacker@europa-uni.de.
[2] University of Ulm, Institute for Applied Analysis, Helmholtzstr. 18, 89081 Ulm, Germany, emil.wiedemann@uni-ulm.de
[3] Max Planck Institute for Software Systems, Campus E1 5, 66123 Saarbrücken, Germany, meikezehlike@mpi-sws.de

## 2.1 Antidiscrimination Law

Concerning antidiscrimination law, it should first be noted that, in our view, it clearly covers instances of automated decision making. The scope of antidiscrimination legislation in the EU is not limited to activities mainly or exclusively undertaken by human beings. Rather, the provisions are formulated in a technologically neutral manner. Perhaps even more importantly, from a teleological perspective, it seems clear that the objective of antidiscrimination law – to prevent material and immaterial harm stemming from discriminatory practices – demands its application to AI systems as well. Generally, intentional practices will qualify as direct discrimination (e.g., masking bias in data sets). By contrast, most instances of discrimination by AI tools will amount to indirect discrimination, particularly if they are a mere side effect of machine learning optimization brought about unintentionally.

Importantly, however, instances of discrimination are not prohibited per se under EU law. Rather, they can generally be justified if the decision maker pursues a legitimate aim and the discriminatory practice is proportionate to that aim. Courts will undertake a comprehensive analysis of all circumstances to evaluate the merits of a claim for justification. As we have explained in greater detail elsewhere [Ha18], however, the chances of justifying algorithmic discrimination are crucially influenced by the source of the bias. If the discriminatory outcome is an artifact of "biased training" (such as biased training data or inadequate feature/target selection), it will be quite hard to justify the outcome. Decision makers must bear reasonable costs to avoid biased training and modeling processes. If, however, bias arises because the targeted qualities are, in reality, unequally distributed between different protected groups, one might speak of 'unequal ground truth'. In these cases, jurisprudence by the Court of Justice of the European Union (CJEU) suggests that it will be easier to meet the burden of justification if the disparate outcome reflects a 'real difference' between the groups that is relevant for the decision at hand.[4]

For potential victims of discrimination, the importance of the source of bias for justification raises an important barrier to justice, however. To differentiate between the different sources, and to gauge potential costs of litigation, plaintiffs would need access to the data and the model. Under discrimination law, however, they generally lack access rights if they merely suspect that the decision might be driven by discrimination.[5] Moreover, they are often not even in a position to establish a prima facie case of statistically significant differential treatment without access to data and the model.

## 2.2. Data Protection Law

At this stage, data protection law, and particularly the GDPR, enter the scene. Not only does it comprise access rights (Art. 15 GDPR), but it also contains public enforcement tools which might remedy the enforcement deficiencies of antidiscrimination law. However, for these GDPR tools to be available, algorithmic discrimination would have to qualify as a breach of data protection law. There are two main routes to bring about this result.

First, Article 5(1)(a) GDPR contains the principles of accuracy and of fair data processing. Arguably, these principles are breached in cases of unjustified discrimination [A18]. This entails an integrated understanding of data protection and antidiscrimination law. Second, in cases of automated decision making, Article 22(3) GDPR mandates that the controllers must take measures to safeguard the rights and freedoms of data subjects. Scholars have rightly argued that bias reduction mechanisms must count among those necessary safeguards [see ref. in Ha18, p. 1177].

Such potential violations of the GDPR open up the toolbox of GDPR enforcement. For example, data protection authorities may impose heavy fines (Art. 83 GDPR), and may perform algorithmic audits (Art. 58(1)(b) GDPR). Hence, the conceptual convergence of antidiscrimination and data protection law facilitates a much more robust enforcement mechanism for cases of AI bias under EU law.

However, even such an integrated understanding of EU law is unlikely to overcome the challenges of algorithmic discrimination on its own. Two main problems remain with this approach. First, data protection authorities, upon which much of the enforcement burden would rest, are notoriously under-resourced and, in some cases, may lack the technical competences to rigorously audit complex AI systems. Second, many of the instruments of data protection law are ex post correction strategies. This means, however, that the harm has already occurred, which may be quite severe in cases of discrimination.

---

4 See, e.g., CJEU, Case C-96/80, *Jenkins*, EU:C:1981:80, para 12; Case C-170/84, *Bilka-Kaufhaus*, EU:C:1986:204, para 36.
5 See CJEU, Case C-415/10, *Meister*, EU:C:2012:217, para. 46.

# 3   Matching Code and Law: Algorithmic Fairness and Optimal Transport

In view of the mentioned shortcomings of a purely legalistic approach, it seems attractive to look for strategies of preventing discrimination ex ante by implementing non-discrimination principles directly at the code level. This is precisely the object of an ever-growing research effort conducted around the world under the banner of algorithmic fairness. While an overview of the many fairness definitions employed in the computer science literature transcends the scope of this paper [DL19; PS20], most definitions can be categorized into two main groups: individual fairness and group fairness [DH+12; FS+16].

### 3.1.   The Divide between Individual and Group Fairness

Individual fairness compares attributes and outcomes for single individuals. More specifically, it usually demands that two individuals that are similar in terms of their attributes (with the exception of protected attributes) are mapped, by the algorithmic process, onto a similar output [DH+12]. This corresponds to the old Aristotelian notion of 'treating likes alike'. Intriguingly, such an understanding also matches the definition of equality before the law (Art. 20 of the Charter of Fundamental Rights of the EU) in the jurisprudence of the CJEU.[6] Group fairness, by contrast, compares outcomes at the group level [DH+12; PS20]. One common metric, statistical parity, demands that the same proportion of individuals must be positively selected from each protected group [YS+18; PS20]. This corresponds to a more outcome-egalitarian concept of social justice [B18]. Moreover, since the groups are treated in a statistically equivalent way, it is very hard to find indirect discrimination once statistical parity is fulfilled.

Importantly, however, there is generally a trade-off between group fairness and individual fairness if the true score distributions between the protected groups differ at the outset [FS+16; ZH+20; but see also B20; ZC20]. In such a case, enforcing group fairness implies that some individuals from the discriminated group will be positively selected while similarly qualified individuals from the privileged group will be rejected, breaching individual fairness. Therefore, the discussion around the correct fairness metrics reproduces long-standing philosophical debates about meritocratic vs. egalitarian concepts of social justice [B18].

### 3.2.   Bridging the Divide

In a model described in greater detail elsewhere [ZH+20], we propose to bridge this development by continuously interpolating between measures of individual and group fairness. To this end, we define a mapping from individuals' raw scores (the outcome of some ML process) to fair scores. Hence, our fairness tool generally functions as a post-processing approach, but it may theoretically also be used as a pre-processing tool by applying it to the target value of the training data points. In the mapping, we introduce a parameter ($\theta$) which allows to fine-tune the degree to which the raw score distributions of different protected groups are approximated. More precisely, we calculate a barycenter of the different group raw score distributions. The barycenter represents an intermediate distribution with minimal distance from the various raw group score distributions in a least square sense.[7] The parameter $\theta$, which runs from 0 to 1, determines the degree to which each raw score distribution is shifted toward the barycenter. If $\theta$ equals 0, the raw score distributions are left unchanged and individual differences between the scored individuals are fully preserved. This minimizes what we call an individual fairness error.[8] In the other extreme, if $\theta$ is set to 1, distributions are fully matched onto the barycenter, achieving statistical parity. While our mapping guarantees monotonicity within groups, the ranking is usually changed, under high $\theta$ values, for individuals who belong to different groups. Hence, the individual fairness error rises. Moreover, to the extent that the raw scores correctly represent the target qualities, the enforcement of group fairness breaches similarity-based definitions of individual fairness if the raw scores were unequally distributed between the different groups.

---

[6] See, e.g., CJEU, Case C-149/10, *Chatzi*, ECLI:EU:C:2010:534, para. 64.
[7] See Theorem 2.4 in [ZH+20] for details.
[8] See Equation 2.9 in [ZH+20] for details.

Importantly, we use optimal transport theory to minimize the information loss of the decision maker in the mapping process. Not only is the barycenter calculated through optimal transport, but the mapping of each individual group toward the barycenter, to the degree defined by θ, also follows optimal transport. Under the assumption that the raw scores correctly reflect the target qualities, the mapping therefore maximizes decision maker utility under varying flexible fairness constraints defined by the choice of the θ value.

### 3.3. Advantages and Limitations of Our Model

Empirical evaluations of the model on synthetic data and on the LSAT data set show that the choice of θ indeed directly determines the mentioned individual and group fairness values. The validation points to significant advantages, but also limitations of our model.

On the positive side, first, the model can match legal standards by choosing θ in such a way that the approximation of the group distributions prevents the finding of indirect discrimination. As mentioned, indirect discrimination presupposes a statistical disparity between the positive selection probabilities of the different groups. Hence, our model can be used as an important building block for a compliance model for AI systems. Second, the brief legal discussion has shown that differential outcomes between protected groups may be justified depending on the circumstances of the case. Our model provides a flexible framework with which decision makers, and regulators, may adapt the outcome of algorithmic processes to various situations in which different degrees of individual or group fairness may be warranted. In each of these situations, optimal transport guarantees that decision maker utility is maximized under the varying fairness constraints. Third, by choosing different θ values for different groups, the model allows to consciously push certain particularly disadvantaged groups. This allows us to handle problems of intersectional discrimination in which certain subgroups are disadvantaged because of multiple protected attributes. Fourth, finally, the model works equally for one-dimensional and multi-dimensional scores. This may be important if a scoring process contains assessments from different sources on different scales.

The use of optimal transport also limits the model to a certain extent, however. First, to function well, it necessitates group sizes of at least several hundred individuals. Hence, the model is particularly well-suited for Big Data applications, but less so for individual scenarios with few candidates. Second, it presupposes that the raw scores, while potentially imperfect, represent a useful approximation of the true target scores. Otherwise, the impact of the mapping on decision maker utility cannot be guaranteed. However, this constraint seems rather weak. If the scoring process results in unreliable raw scores, the developer must modify the scoring procedure. To the extent that the decision maker intends to use the scores, the assumption that the scores are meaningful does not seem far-fetched.

## 4 Normative Challenges for Algorithmic Fairness

Technical solutions can help to mitigate bias in decision making. The possibility to consciously choose certain fairness parameters constitutes a significant advantage of AI versus human decision making. Nevertheless, important normative challenges remain.

### 4.1. The Choice of θ

In the context of our model, one obvious question relates to the choice of θ. Institutionally, decision makers could be granted leeway to choose θ as they please (within the constraints of antidiscrimination law). However, in certain areas of pronounced societal importance (such as education, housing etc.) one may imagine that the legislator, regulatory agencies, or the courts define specific thresholds or even concrete values for the choice of θ [see also VB17].

Overall, that choice will reflect the trade-off between more individualistic, meritocratic approaches to social justice on the one hand and more outcome-egalitarian ones on the other. In our view, one possibility might be to draw on a capabilitarian approach [S09]. From this perspective, a low θ value could be chosen if two conditions are cumulatively met [ZH+20]. First, there is high confidence in the correctness of the raw scores. Second, a meritocratic allocation regime is normatively desired because the decision does not affect basic capabilities or resources. Conversely, a higher θ value could be selected if decision makers only have an intermediate confidence in the correctness of the raw scores i.e., the scores have some information utility for the decision maker, but may to a certain extent be affected by bias (e.g., biased training). A higher θ value would preserve the in-group rankings, and hence the information encoded in the scores, while mitigating the effect of bias between the groups. Moreover, even if the raw scores are likely to be correct, high θ values might be justified for normative reasons if the allocation of basic resources and capabilities is at stake.

### 4.2. Legal Constraints

Finally, the enforcement of group fairness may engender conflicts with affirmative action law [Ha18; He20; Be20; ZH+20]. In the EU, as in other jurisdictions, it is difficult to draw the line between remedying unjustified discrimination and breaching the right to equality of those individuals negatively affected by affirmative action. This trade-off precisely mirrors the fundamental difference between individual and group fairness discussed above. In our view, under EU law, it must generally be possible to modify a selection procedure, including the raw scores, if the certification of the procedure would lead to unjustified discrimination (e.g., biased training). Otherwise, the restraints of affirmative action law would force the decision maker to violate basic non-discrimination standards.

What remains doubtful, however, is the extent to which decision makers can engage in affirmative action if the original outcome or procedure would have been legally justified (e.g., unequal ground truth). In the EU, standards differ based on whether an affirmative action policy (like the fairness intervention) is enacted before or after a first ranking of the candidates has been conducted. While criteria are more lenient before that first selection (Badeck case),[9] the jurisprudence of the CJEU is quite restrictive after it. In the Marschall case, the Court ruled that a re-ranking based on protected attributes must not be absolute and unconditional.[10] Rather, it may only be undertaken on the basis of an objective assessment taking into account all specific criteria of the affected individuals. In this way, the Court seeks to ensure that individual characteristics speaking against downgrading an applicant (e.g., specific individual hardship for single parents) are not blindly overridden by affirmative action preference.

Overall, the distinction between the Badeck and the Marschall case suggests a dividing line between post-processing approaches on the one hand and pre-/in-processing approaches on the other: the latter are easier to justify, under EU affirmative action law, than the former. Similarly, under US law, re-ranking based on 'race-norming' individual test results is prohibited in employment contexts [GJ96], making post-processing approaches difficult to justify in this domain [RB+20]. More generally, in the Ricci case,[11] the US Supreme Court seemed to adopt a more lenient stance toward interventions at the test-design stage, rather than post-processing modifications [KH+17; K17; He20].

It seems unclear, however, whether such a distinction between the different interventions is normatively and legally justified. For a start, post-processing approaches afford the advantage that the re-ranking result is precisely defined, while in- and pre-processing approaches risk "overshooting", which would be worse for the privileged groups. In our view, it might be worth considering an 'attenuated Marschall standard', under which algorithmic affirmative action is possible, irrespective of the stage of the intervention, as long as meaningful human scrutiny is applied. However, such scrutiny need not accompany each and every case of re-ranking, but might be restricted to those of specific legal interest (e.g., particular individual hardship). In this way, the consideration of important individual criteria in specific cases can be combined with a broad application of algorithmic fairness to large data sets in which human scrutiny of each individual re-ranking decision would often be prohibitively costly.

## 5 Conclusion

Despite significant research efforts in law and computer science over the last years, a consensus on fairness metrics for the purposes of preventing discriminatory outcomes in machine learning contexts has not emerged yet. Arguably, this testifies both to the vagueness of legal standards and to the significantly diverging factual circumstances of the various areas in which AI models are deployed. Therefore, it seems important to develop technical tools that match the necessary openness and flexibility of the legal provisions and the facts of the case. The model described in this paper seeks to contribute to this endeavor.

---

9 CJEU, Case C-158/97, *Badeck*, EU:C:2000:163, paras. 55 and 63 (concerning selection for training and interview).
10 CJEU, Case C-409/95, Marschall, EU:C:1997:533, para. 33.
11 Ricci *v.* DeStefano, 557 U.S. 557, 563 (2009).